\documentclass[conference]{IEEEtran}
\usepackage{amsmath,graphicx}
\usepackage{amsfonts}
\usepackage{enumerate}
\usepackage{bm}
\usepackage{algorithm} 
\usepackage{algorithmic} 

\usepackage{mathrsfs}
\usepackage{stfloats}
\usepackage{enumerate}
\usepackage{epstopdf}
\usepackage{subfigure}
\usepackage{float}
\usepackage{amsbsy}
\usepackage{amsmath}
\usepackage{amssymb}
\usepackage{color}
\usepackage{ntheorem}
\usepackage{cite}

\theorembodyfont{\upshape}
\theoremheaderfont{\rmfamily\itshape}
\theoremseparator{:}
\theoremstyle{remark}

\newtheorem{theo}{\hspace{1em}Theorem}

\newtheorem{corollary}{\hspace{1em}Corollary}
\newtheorem*{pproof}{\hspace{2em}Proof}

\begin{document}
\IEEEoverridecommandlockouts
\title{{Near-Field Beam Focusing Pattern and Grating Lobe Characterization for
 Modular XL-Array}\vspace{-1pt}}
\author{
 \IEEEauthorblockN{Xinrui Li\IEEEauthorrefmark{1}, Zhenjun Dong\IEEEauthorrefmark{1}, Yong Zeng\IEEEauthorrefmark{1}\IEEEauthorrefmark{2}, Shi Jin\IEEEauthorrefmark{1}, Rui Zhang\IEEEauthorrefmark{3}\IEEEauthorrefmark{4}}

 \IEEEauthorblockA{\IEEEauthorrefmark{1}National Mobile Communications Research Laboratory,
 and Frontiers Science Center for Mobile Information Communication \\and Security,
 Southeast University, Nanjing 210096, China}

 \IEEEauthorblockA{\IEEEauthorrefmark{2}Purple Mountain Laboratories, Nanjing 211111, China}

  \IEEEauthorblockA{\IEEEauthorrefmark{3}The Chinese University of Hong Kong (Shenzhen), and Shenzhen Research Institute of Big Data, Shenzhen 518172, China
  }
 \IEEEauthorblockA{\IEEEauthorrefmark{4}Department of Electrical and Computer Engineering, National University of Singapore, Singapore 117583, Singapore
 \\Email: \{230218659, zhenjun\_dong, yong\_zeng, jinshi\}@seu.edu.cn, rzhang@cuhk.edu.cn}

}
\maketitle
\vspace{-1cm}

\begin{abstract}
In this paper, we investigate the near-field modelling and analyze the beam focusing pattern for modular extremely large-scale array (XL-array) communications. As modular XL-array is physically and electrically large in general, the accurate characterization of amplitude and phase variations across its array elements requires the non-uniform spherical wave (NUSW) model, which, however, is difficult for performance analysis and optimization. To address this issue, we first present two ways to simplify the NUSW model by exploiting the unique regular structure of modular XL-array, termed \textit{sub-array based uniform spherical wave (USW) models with different} or \textit{common angles}, respectively. Based on the developed models, the near-field beam focusing patterns of XL-array communications are derived. It is revealed that compared to the existing collocated XL-array with the same number of array elements, modular XL-array can significantly enhance the spatial resolution, but at the cost of generating undesired grating lobes. Fortunately, different from the conventional far-field uniform plane wave (UPW) model, the near-field USW model for modular XL-array exhibits a higher grating lobe suppression capability, thanks to the non-linear phase variations across the array elements. Finally, simulation results are provided to verify the near-field beam focusing pattern and grating lobe characteristics of modular XL-array.
\end{abstract}
\section{Introduction}
Massive multiple-input and multiple-output (MIMO) by using large antenna arrays has greatly enhanced the wireless communication spatial resolution and spectral efficiency, thus regarded as  a key enabling technology for the fifth-generation (5G) wireless network. In the future sixth-generation (6G) network, the continuous growth in number of antennas will render the large antenna array today likely to evolve into the extremely large-scale array (XL-array) \cite{Bjornson2019,  Lu2021}.
In particular, modular XL-array is a promising new array architecture, where array elements are regularly arranged on a common platform in a modular manner with the adjacent
modules separated by a given distance, so as to cater to different mounting structures \cite{Bertilsson2018, Li2022, Li2023}. For example, modular XL-array can be installed on building facades with the array modules separated by windows \cite{Li2022, Li2023}. Compared to the existing collocated XL-array with a single module only \cite{Zeng2021, Lu2021}, modular XL-array can not only achieve more flexible and conformal deployment, but also enhance the spatial resolution due to increased array aperture.\par

The large aperture of modular XL-array renders the conventional far-field channel modelling based on uniform plane wave (UPW) assumption practically invalid \cite{Zeng2021,122}. In fact, compared with its collocated counterpart, the physical dimension of modular XL-array needs to take into account the module separations, which makes users or scatterers more likely being located in its near-field region \cite{Li2022, Li2023}. As such, instead of the widely used far-field UPW model, the accurate characterization of the amplitude and phase variations across modular XL-array elements requires the more complicated non-uniform spherical wave (NUSW) model, which has received growing attention recently \cite{Zeng2021,Lu2021,Dong2022}. For example, the authors in \cite{Lu2021} and \cite{Zeng2021} developed the NUSW-based near-field modelling for collocated XL-array communications, and further derived the closed-form expression of the achievable receiver signal-to-noise ratio (SNR). The NUSW-based near-field spatial correlation in non-line-of-sight (NLoS) environment was derived in \cite{Dong2022}, which revealed the spatially wide-sense non-stationary characteristic. However, the aforementioned works only studied the near-field modelling and performance analysis for collocated XL-array, which in general cannot apply to modular XL-array. In our previous works \cite{Li2022, Li2023}, wireless communication with modular XL-array was investigated, for which closed-form expression for the achievable SNR under NUSW modelling was derived. However, SNR expression only gives the performance metric for single-user system. For the more general multi-user communication systems with inter-user interference (IUI), other properties like the beam focusing pattern of modular XL-array are crucial to characterize the user performance. This thus motivates our current work.\par

In this paper, we investigate the near-field modelling and provide the beam focusing pattern analysis for modular XL-array communications. First, by exploiting the unique regular structure of modular XL-array, we introduce two practical ways to simplify the NUSW-based near-field model, termed \textit{sub-array based USW models with different} or \textit{common angles}, respectively. In particular, with the latter model, we show that the near-field array response vector for \textit{modular XL-uniform linear array} (mXL-ULA) can be represented as a Kronecker product of the array response vectors of  a sparse ULA and a collocated ULA. Based on the developed models, we analyze the near-field beam focusing patterns of mXL-ULA. It is revealed that compared to the conventional collocated XL-ULA with the same number of array elements, mXL-ULA can significantly enhance the spatial resolution, but at the cost of generating undesired grating lobes. Fortunately, different from the conventional far-field UPW model, the near-field USW model enables mXL-ULA to exhibit a higher grating lobe suppression capability, thanks to our accurate characterization of the non-linear phase variations across the array elements of different modules. Finally, simulation results are provided to demonstrate the near-field beam focusing pattern and grating lobe characteristics of modular XL-array communications. \par

It is worth noting that grating lobes are well understood under the conventional far-field UPW model for sparse arrays, where inter-element separation is larger than half of the wavelength \cite{Zhuang2008,  Gao2022, krivosheev2010}. Research efforts have also been devoted to develop effective techniques for grating lobe suppression, such as frequency division, optimization of inter-element spacing and phase shifts \cite{Zhuang2008, krivosheev2010}. However, prior studies mainly considered the far-field UPW assumption for sparse arrays used in e.g. radar applications. Thus, their results cannot be applied to the near-field modular XL-array modelling studied in this paper. \par

\section{System model} \label{model}\vspace{-1pt}
As shown in Fig. 1, we consider a wireless communication system
where the base station (BS) is equipped with an mXL-ULA.
The total number of array elements is $N M$, where $N$ is the number of modules and $M$ is the number of antenna elements in each module.
Let $d$ represent the inter-element spacing for antennas in each module, which is typically on the wavelength scale, e.g., $d=\frac{\lambda}{2}$, with $\lambda$ denoting the signal wavelength. Therefore, the sub-array of each module has the physical size $S=(M-1)d$. Furthermore, let the separation between the reference elements (say the center elements) of adjacent modules be denoted as
$\Gamma d$, where $\Gamma\geq M$ is a parameter that depends on the size of the discontinuity of the actual mounting structure. For example, for modular XL-array mounted on building facades separated by windows, $\Gamma$ depends on the window size.
The total physical size
of the mXL-ULA is thus $D=[(N-1)\Gamma+(M-1)]d$. In particular, when $\Gamma=M$, such a modular XL-array architecture degenerates to the conventional collocated XL-array \cite{Lu2021,Zeng2021}. For ease of notations, $N$ and $M$ are assumed to be odd numbers,
so that the module index $n$ and the antenna index $m$ for each module belong to the integer sets $\mathcal N= \left\{0,\pm 1,\cdots,\pm \frac{N-1}{2}\right\}$ and $\mathcal M=\left\{0, \pm 1,\cdots,\pm \frac{M-1}{2}\right\}$,  respectively.
Without loss of generality, we assume that the mXL-ULA is placed along the $\emph{y}$-axis, and it is symmetric about the origin. Therefore, the location of the $m$-th element within module $n$ is
${\bf w}_{n,m}=\left[0,y_{n,m}\right]^T$, where $y_{n,m}=(n\Gamma+m)d$, $\forall n\in\mathcal N$, and $\forall m \in \mathcal M$.

Consider a user or a scatterer at location ${\bf q}=[r\cos\theta, r\sin\theta]^T$, where $\theta \in [-\frac{\pi}{2},\frac{\pi}{2}]$ denotes the angle with respect to the positive $x$-axis, and $r$ represents its distance from the array center.
The distance between $\mathbf q$ and the $m$-th array element of module $n$ is given by
\begin{equation}\label{EQU-1} \vspace{-3pt}
\begin{split}
r_{n,m}&=||{\bf q}-{\bf w}_{n,m}||\\
&=\sqrt{r^2-2r y_{n,m}\sin{\theta}+y_{n,m}^2},\forall n\in\mathcal N, m\in\mathcal M.\\
\end{split}
\end{equation}\par
As such, to accurately characterize the signal amplitude and phase variations across the $NM$ array elements, the NUSW-based near-field channel vector of the mXL-ULA for a user/scatterer with distance $r$ and angle $\theta$, denoted as ${\bf h}(r,\theta)\in \mathbb{C}^{(N M)\times 1}$, can be modelled as \cite{Lu2021}
\begin{equation}\label{EQU-21} \vspace{-3pt}
\begin{split}
{\bf h}(r,\theta)&=\left[\frac{\sqrt{\beta_0}}{r_{n,m}}e^{-j\frac{2\pi}{\lambda}r_{n,m}}\right]_{n\in\mathcal N,m\in\mathcal M}=\frac{\sqrt{\beta_0}}{r}{\bf a}(r,\theta),\\
\end{split}
\end{equation}
where $\beta_0$ is the reference channel power gain at the distance of $1$ meter ($\rm m$), and ${\bf a}(r,\theta)=\left[\frac{r}{r_{n,m}}e^{-j\frac{2\pi}{\lambda}r_{n,m}}\right]_{n\in\mathcal N,m\in\mathcal M}$ is the array response vector for mXL-ULA under the NUSW model \cite{Zeng2021, Li2022}. When $r\ge 1.2D$, it has been shown that the amplitude variations across array elements can be neglected \cite{122}. In this case, the NUSW model degenerates to the USW model, under which the array response vector ${\bf a}(r,\theta)$ simplifies as
\begin{equation}\label{EQU-21} \vspace{-3pt}
\begin{split}
&{\bf a}(r,\theta)=\left[e^{-j\frac{2\pi}{\lambda}r_{n,m}}\right]_{n\in\mathcal N,m\in\mathcal M}.\\
\end{split}
\end{equation}\par
\begin{figure}[t]  \vspace{-0.14cm}
  \centering
    \includegraphics[scale=0.66]{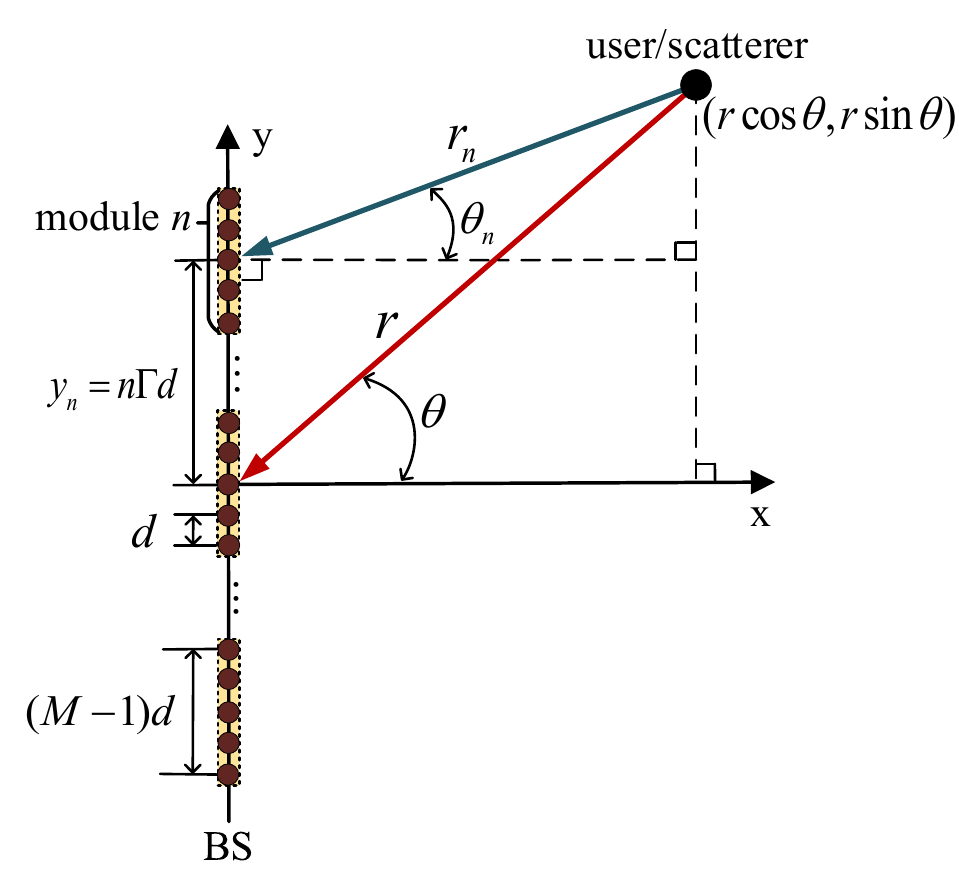}
     \vspace{-0.3cm}
  \caption{An mXL-ULA with $N$ modules and $M$ antennas in each module. }\label{pic1}
  \label{12}
  \vspace{-0.5cm}
\end{figure}\par

However, (3) is still difficult for performance analysis since it depends on the sophisticated distance expression (1). To obtain more tractable near-field models, we propose two simplified versions of (3) by exploiting the unique regular structure of mXL-ULA, which we term as \textit{sub-array based USW model with different} or \textit{common angles}, respectively. Before that, we first present the conventional UPW-based far-field model for modular arrays as a benchmark. \par
\textit{1) Conventional UPW model}:  When $r\ge\frac{2D^2}{\lambda}$, the user/scatterer is located in the far-field region of the whole array. In this case, by applying the first-order Taylor approximation to the distance expression in (1), we have $r_{n,m}\approx r-y_{n,m}\sin\theta $.  As a result, the array response vector ${\bf a}(r,\theta)$ in (3) degenerates to the existing UPW model, i.e.,
\begin{equation}\label{EQU-21} \vspace{-3pt}
\begin{split}
&{\bf a}(r,\theta)=e^{-j \frac{2\pi}{\lambda}{r}}{\bf p}(\theta)\otimes {\bf b}(\theta),\\
\end{split}
\end{equation}
where ${\bf p}(\theta)=\left[e^{j \frac{2\pi}{\lambda}{n\Gamma d\sin\theta}}\right]_{n\in\mathcal N}$ and ${\bf b}(\theta)=\left[e^{j \frac{2\pi}{\lambda}{md\sin\theta}}\right]_{m\in\mathcal M}$.
Note that ${\bf a}(r,\theta)$ in (4) is expressed as a Kronecker product of the UPW-based array response vectors of an $N$-element sparse ULA with element separation $\Gamma d$, i.e., ${\bf p}(\theta)$, and an $M$-element collocated ULA with element separation $d$, i.e., ${\bf b}(\theta)$. Fig. 2 illustrates the geometric relationship of three array architectures, together with their respective array response vectors.  \par
\textit{2) Sub-array based USW model with different angles}: When $\frac{2S^2}{\lambda}\le r<\frac{2D^2}{\lambda}$, the user/scatterer is located in the far-field region of each array module, but in the near-field region of the whole array. In this case, the array response vector ${\rm a}(r,\theta)$ in (3) can be simplified by applying the UPW model within each module $n$, while the near-field spherical property is reflected by waves across modules.
To this end, by letting $m=0$ in (1),
$r_{n}\triangleq r_{n,0}=\sqrt{r^2-2r y_{n}\sin{\theta}+y_n^2}, \forall n\in\mathcal N$, denotes the distance between $\mathbf q$ and the reference element of module $n$, with $y_n=y_{n,0}=n\Gamma d$.
Besides, let $\theta_n\in [-\frac{\pi}{2},\frac{\pi}{2}]$ denote the direction of user/scatterer $\mathbf q$ with respect to module $n$. It follows from Fig. 1 that we have
$\sin\theta_n=\frac{r\sin\theta-y_n}{r_n}, \forall n\in\mathcal N$.
Hence, the array response vector ${\bf a}(r,\theta)$ in (3) simplifies as
\begin{equation}\label{EQU-21} \vspace{-3pt}
\begin{split}
&{\bf a}(r,\theta)=\left[e^{-j\frac{2\pi}{\lambda}r_n}{\bf b}(\theta_n)\right]_{n\in\mathcal N},\\
\end{split}
\end{equation}
where ${\bf b}(\theta_n)=\left[e^{j\frac{2\pi}{\lambda}md\sin\theta_{n}}\right]_{m\in\mathcal M}$.
Compared with the UPW model in (4), the near-field effect of ${\bf a}(r,\theta)$ in (5) is reflected in two aspects. Firstly, the phase variations across modules are non-linear, since  the exact distance $r_n$, instead of its first-order Taylor approximation, is used to model the phase of the reference elements of each module. Secondly, the angle of arrival/departure (AoA/AoD) varies across different modules, since $\theta_n$ depends on the module index $n$.\par
\textit{3) Sub-array based USW model with common angle}: When $\max\{5D, \frac{4SD}{\lambda}\}\le r< \frac{2D^2}{\lambda}$, we consider that the user or scatterer is located in the extended  far-field region of each
module, while in the near-field region of the whole array.  The so-called extended far-field region is specified by the new distance criterion $\max\{5D, \frac{4SD}{\lambda}\}$, which is guaranteed to be larger than the Rayleigh distance of each module, $\frac{2S^2}{\lambda}$, due to $D>\frac{S}{2}$. The relevant proof is given in Appendix A. In this case, the angles $\theta_n$ viewed from different modules are approximately equal, i.e., $\theta_n \approx \theta, \forall n\in\mathcal N$.  Therefore, the array response vector ${\bf a}(r,\theta)$ in (5) further simplifies as
\begin{equation}\label{EQU-21} \vspace{-3pt}
\begin{split}
&{\bf a}(r,\theta)={\bf e}(r,\theta)\otimes {\bf b}(\theta),\\
\end{split}
\end{equation}
where
${\bf e}(r, \theta)=\left[e^{-j\frac{2\pi}{\lambda}r_{n}}\right]_{n\in\mathcal N}$.
Different from the UPW model in (4), ${\bf e}(r,\theta)$ in (6) is the USW-based near-field array response vector of an $N$-element sparse ULA with adjacent elements separated by $\Gamma d$. Similar to (4), the array geometric relationship corresponding to (6) is also shown in Fig. 2.\par
\begin{figure}[htbp]
\centering
\includegraphics[scale=0.6]{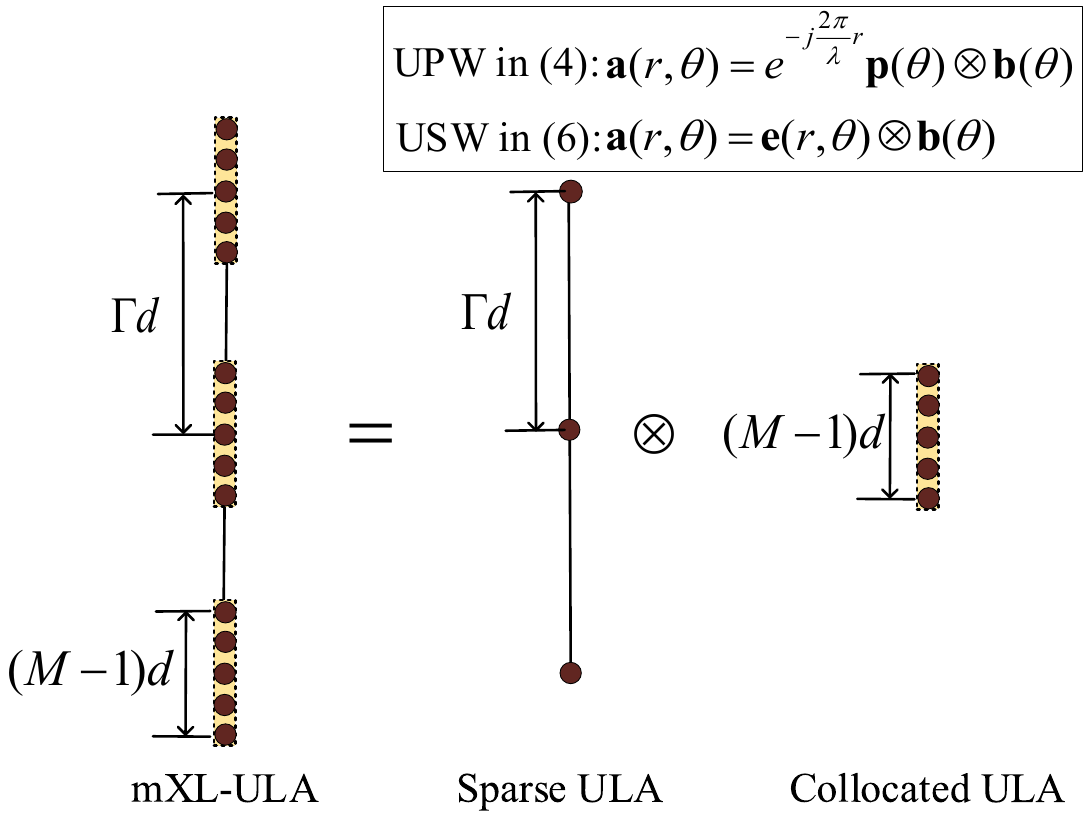} \vspace{-0.2cm}
\caption{The equivalent geometry of mXL-ULA under UPW model and USW model with common angle, which can be interpreted as the Kronecker product of an $N$-element sparse ULA and an $M$-element collocated ULA.}\label{movement}
  \vspace{-0.5cm}
\end{figure}

\section{Near-field beam focusing pattern and grating lobes of mXL-ULA} \label{model}\vspace{-1pt}
Based on the models presented in Section II, we study the near-field beam focusing patterns of mXL-ULA in this section.
For the conventional UPW-based far-field model, beam pattern tells the intensity variation of a designed beam intended for a certain direction as a function of the actual observation direction.
In contrast, for the near-field model, we use the term \textit{beam focusing pattern}, since instead of the directions, it is the actual locations that completely specify the signal strength variations.
Specifically, for a near-field beam designed for the intended location $(r',\theta')$, the beam focusing pattern at the observation location $(r,\theta)$ can be defined as
\begin{equation}\label{EQU-21} \vspace{-3pt}
\begin{split}
G(r,\theta;r',\theta')&\triangleq \frac{1}{MN}{\left|{\bf a}(r',\theta')^H{\bf a}(r,\theta)\right|}\\
&=\frac{1}{MN}{\left|\sum_{n\in\mathcal N}\sum_{m\in\mathcal M}e^{-j\frac{2\pi}{\lambda}(r_{n,m}-r'_{n,m})}\right|},\\
\end{split}
\end{equation}
where we have used the USW-based near-field array response vector in (3). However, the direct beam pattern analysis based on (3) is quite challenging. To obtain more insights, we provide the analysis of beam patterns of mXL-ULA for the three simplified models presented in Section II.
\subsection{Far-Field Beam Pattern under the UPW Model.} \label{model}\vspace{-1pt}

\begin{theo}\label{theo1}
The far-field beam pattern of modular ULA under the UPW model can be expressed as
\begin{equation}\label{EQU-10} \vspace{-3pt}
\begin{split}
G(\theta;\theta')
=\underbrace{\left|\frac{\sin \left(\pi N \Gamma \bar{d}\Delta_\theta\right)}{N\sin \left(\pi\Gamma \bar{d}\Delta_\theta\right)}\right|}_{\left|H_{N,\Gamma\bar{d}}(\Delta_\theta)\right|}\underbrace{\left|\frac{\sin \left(\pi M \bar{d}\Delta_\theta\right)}{M\sin \left(\pi \bar{d} \Delta_\theta\right)}\right|}_{\left|H_{M,\bar{d}}(\Delta_\theta)\right|},\\
\end{split}
\end{equation}
where $\Delta_\theta=\sin\theta-\sin\theta'$, ${\bar d}=\frac{d}{\lambda}$ is the inter-element spacing normalized by wavelength, and $H_{\tilde{M},\tilde{d}}(\Delta_\theta)=\frac{\sin \left(\pi \tilde{M}\tilde{d}\Delta_\theta\right)}{\tilde M\sin \left(\pi\tilde{d}\Delta_\theta\right)}$ is the Dirichlet kernel function.
\begin{IEEEproof}
Theorem 1 can be readily proved by substituting (4) into (7).
\end{IEEEproof}
\end{theo}\par

Theorem 1 shows that under the far-field UPW model, the beam pattern $G(\theta;\theta')$ only depends on $\Delta_\theta$, i.e., the difference of the sine of the angles, also known as the spatial frequencies.  Besides, the beam pattern is given by the product of the beam pattern $\left|H_{N,\Gamma\bar{d}}(\Delta_\theta)\right|$ of the $N$-element sparse ULA with element separation $\Gamma d$, and the beam pattern $\left|H_{M,\bar{d}}(\Delta_\theta)\right|$ of the $M$-element collocated ULA with element separation $d$.
To gain further insights, some important properties of the function   $|H_{\tilde{M},\tilde{d}}(\Delta_\theta)|$ are given below.\par

{\it 1) Angular Resolution:} By letting  $\pi \tilde{M}\tilde{d} \Delta_\theta=\pm \pi$ so that
$H_{\tilde M, \tilde d}(\Delta_\theta)=0$,
we can obtain the null-to-null beam width of the main lobe of $H_{\tilde M, \tilde d}(\Delta_\theta)$, i.e.,
${\rm BW}=\frac{2}{\tilde{M}\tilde{d}}$ \cite{222}.
Therefore, the angular resolution of the two terms in (8) are $\frac{2}{\Gamma N\bar{d}}$ and $\frac{2}{M\bar{d}}$, respectively. Since $\Gamma N>M$ in general, the overall spatial resolution is given by the finer value  $\frac{2}{\Gamma N\bar{d}}$, which depends on the total array aperture.\par
{\it 2) Grating Lobes:} For $H_{\tilde{M},\tilde{d}}(\Delta_\theta)$, when the inter-element spacing is larger than half of the wavelength, i.e., $\tilde{d}>\frac{1}{2}$, grating lobes that have equal amplitude as the main lobe will appear. By letting $\pi \tilde{d}\Delta_\theta= k\pi$, $k=\pm1,\pm2,\cdots$, we can obtain the angular direction of each grating lobe, i.e.,
$\Delta_\theta= k \frac{1}{\tilde{d}}, k=\pm1,\pm2,\cdots$\cite{222}.
For the considered modular ULA, since $\Gamma \bar{d}>\frac{1}{2}$ in general, the term $\left|H_{N,\Gamma\bar{d}}(\Delta_\theta)\right|$ in (8) will lead to the appearance of grating lobes that are separated by  $\frac{1}{\Gamma\bar{d}}$, which causes angular ambiguity.

\begin{corollary}\label{lemma2}
By substituting $\Gamma=M$,  $G(\theta;\theta')$ in (8) reduces to that of the conventional collocated ULA \cite{2022}, i.e.,
\begin{equation}\label{EQU-6} \vspace{-3pt}
\begin{split}
G(\theta;\theta')=\underbrace{\left|\frac{\sin \left(\pi MN \bar{d}\Delta_\theta\right)}{MN\sin \left(\pi \bar{d}\Delta_\theta\right)}\right|}_{\left|H_{MN,\bar{d}}(\Delta_\theta)\right|}.\\
\end{split}
\end{equation}
\end{corollary}\par
Thus, the angular resolution of the collocated ULA in (9) is $\frac{2}{MN\bar{d}}$, which is worse than $\frac{2}{\Gamma N\bar d}$ of the modular counterpart, since $\Gamma\ge M$. This shows the superiority of modular array under the UPW model for improving the angular resolution, but at the cost of generating grating lobes. \par


\subsection{Beam Focusing Pattern under the Sub-Array Based USW Model with Different Angles.} \label{model}\vspace{-1pt}

\begin{theo}\label{theo2}
The near-field beam focusing pattern of mXL-ULA under the sub-array based USW model with different angles can be expressed as
\begin{equation}\label{EQU-21}  \vspace{-0.2cm}
\begin{split}
&G(r,\theta;r',\theta')=\frac{1}{N}\Bigg |\sum_{n\in\mathcal N}  e^{-j\frac{2\pi}{\lambda}\Delta_{r,n}}\underbrace{\frac{\sin \left(\pi M \bar{d}\Delta_{\theta,n}\right)}{M\sin \left(\pi \bar{d} \Delta_{\theta,n}\right)}}_{H_{M,\bar{d}}(\Delta_{\theta,n})}\Bigg |,\\
\end{split}
\end{equation}
where $\Delta_{r,n}=r_n-r_n'$ and $\Delta_{\theta,n}=\sin\theta_n-\sin\theta_n', \forall n\in\mathcal N$.
\begin{IEEEproof}
Theorem 2 can be readily shown by substituting (5) into (7).
\end{IEEEproof}
\end{theo}\par
Theorem 2 shows that the near-field beam focusing pattern of mXL-ULA is given by a weighted sum of $N$ beam patterns, each of which corresponds to one module that has different spatial frequency difference $\Delta_{\theta,n}$.  The complex-valued weighting coefficient $e^{-j\frac{2\pi}{\lambda}\Delta_{r,n}}$ depends on the exact phase difference at the reference elements of each module. It can be shown that the near-field expression (10) includes the far-field result (8) as a special case when $r\geq \frac{2D^2}{\lambda}$.
However, it is still difficult to obtain a closed-form expression for (10), since $\Delta_{r,n}$ and $\Delta_{\theta,n}$ are related to the module index $n$ in a sophisticated manner. In order to get a closed-form expression for (10), we consider the sub-array based USW model with common angle in the following.
\subsection{Beam Focusing Pattern under the Sub-Array Based USW Model with Common Angle.} \label{model}\vspace{-1pt}

\begin{theo}\label{theo2}
The near-field beam focusing pattern of mXL-ULA under the sub-array based USW model with common angle can be expressed as
\vspace{-0.1cm}
\begin{equation}\label{EQU-21} \vspace{-3pt}
\begin{split}
G(r,\theta;r',\theta')=\frac{1}{N}\left|\sum_{n\in\mathcal N} e^{-j\frac{2\pi}{\lambda}\Delta_{r,n}}\right|\left|H_{M,\bar{d}}(\Delta_\theta)\right|,\\
\end{split}
\end{equation}
\begin{IEEEproof}
Theorem 3 directly follows from (10) by letting $\sin \theta_n \approx \sin\theta$ and $\sin\theta_n'\approx \sin\theta'$, $\forall n \in \mathcal N$.
\end{IEEEproof}
\end{theo}\par
Theorem 3 shows that under the sub-array based USW model with common angle, the near-field beam focusing pattern can be split into two separate parts.
The first part is the USW beam focusing pattern of the sparse ULA with $N$ elements, and the second one is the UPW beam pattern of the
collocated ULA with $M$ elements.
When the number of modules $N$ is large,  a closed-form expression of (11) can be obtained, as given below.
\begin{corollary}\label{theo2}
When $N$ is large, $G(r,\theta;r',\theta')$ in (11) can be obtained in closed-form in (12) shown at the top of the next page,
\begin{figure*}
 \vspace{-0.1cm}
\begin{equation}\vspace{-3pt}
\label{eq6}
G(r,\theta;r',\theta')\!=\!\left\{
\begin{aligned}
&\!\frac{\left|\left[F\left(\sqrt{| \nu_{r,\theta}|}\frac{N}{2}+\frac{\mu_{\theta}}{2\sqrt{| \nu_{r,\theta}|}}\right)+F\left(\sqrt{|\nu_{r,\theta}|}\frac{N}{2}-\frac{\mu_{\theta}}{2\sqrt{|\nu_{r,\theta}|}}\right)\right]\right|}{\sqrt{|\nu_{r,\theta}|}N}{\left|H_{M,\bar{d}}(\Delta_\theta)\right|} & , \quad \! \frac{\cos^2\theta}{r}\neq \frac{\cos^2\theta'}{r'}, \\
&{\left|H_{N,\Gamma\bar{d}}(\Delta_\theta)\right|}{\left|H_{M,\bar{d}}(\Delta_\theta)\right|}
& , \quad \!  \frac{\cos^2\theta}{r}= \frac{\cos^2\theta'}{r'},\\
\end{aligned}
\right.
\end{equation}
{\noindent} \rule[0pt]{17.5cm}{0.05em}
  \vspace{-0.6cm}
\end{figure*}
where  $\nu_{r,\theta}=-{\pi\bar d}\Gamma^2 d\delta_{r,\theta}$, $\mu_\theta={2\pi \bar d}\Gamma\Delta_{\theta}$, and $\delta_{r,\theta}=\frac{\cos^2\theta}{r}-\frac{\cos^2\theta'}{r'}$. Besides, $F(x)=C(x)+jS(x)$, with $C(x)= \int_{0}^{x} \cos t^{2} \mathrm{~d} t$ and $S(x)=\int_{0}^{x} \sin t^{2} \mathrm{~d} t$ being Fresnel integrals \cite{Gradshteyn2007}.
\end{corollary}\par
\begin{pproof}Please refer to Appendix B. $\hfill \blacksquare$\end{pproof}  \par

Corollary 2 shows  that the result in (12) is dependent on the so-called angle-distance difference $\delta_{r,\theta}$, and spatial frequency difference  $\Delta_\theta$. Define the curve $r=\xi\cos^2\theta$ in polar domain by using the distance ring $\xi$ \cite{Dai2022}. Interestingly, when two locations $(r,\theta)$ and $(r',\theta')$ are at the same distance ring $\xi$, i.e., $\frac{1}{\xi}=\frac{\cos^2\theta}{r}=\frac{\cos^2\theta'}{r'}$, the USW-based  result in (12) is identical to the UPW-based result in (8). For the general scenario when the two locations are in different distance rings, the USW beam pattern depends on the function $F(x)$ given by Fresnel integrals.
Furthermore, when the two locations are along the same direction but having different distances, i.e., $\Delta_\theta=0$ and $\Delta_r=r-r'\neq 0$, we have $\mu_\theta=0$ and $\nu_r=-{\pi\bar d}\Gamma^2 d\big(\frac{\cos^2\theta'}{r}-\frac{\cos^2\theta'}{r'}\big)$. In this case, the result in (12) reduces to $G_0(r;r')=\frac{\left|F\left(\sqrt{|\nu_r|}\frac{N}{2}\right)\right|}{\sqrt{|\nu_r|}\frac{N}{2}}$.  As a comparison, the result for the conventional collocated counterpart is obtained by letting $\Gamma=M$, giving by $G_1(r;r')=\frac{\left|F\left(\sqrt{|\bar \nu_r|}\frac{N}{2}\right)\right|}{\sqrt{|\bar\nu_r|}\frac{N}{2}}$, where $\bar\nu_r=-{\pi\bar d}M^2 d\big(\frac{\cos^2\theta'}{r}-\frac{\cos^2\theta'}{r'}\big)$.
Since $\Gamma\ge M$ and $\frac{\left|F(x)\right|}{x}$ with $x>0$ is a function that ripplingly decreases with $x$ \cite{Dai2022}, we typically have $G_0(r;r')\le G_1(r;r')$. This indicates that benefiting from its better distance resolution, compared to the collocated counterpart, modular XL-ULA can more effectively mitigate the IUI among users along the same direction.\par
\begin{figure}
\centering
\includegraphics[scale=0.25]{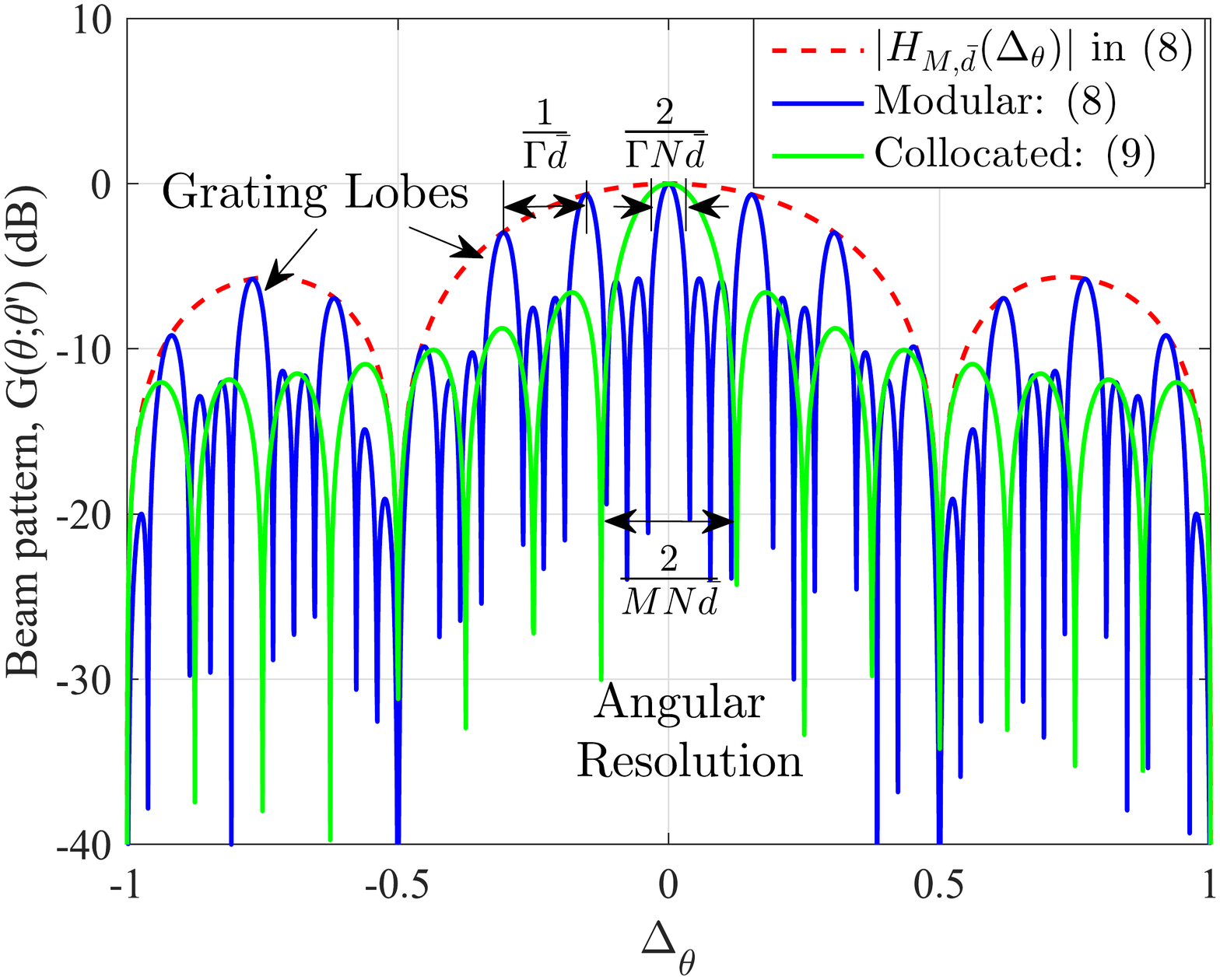} \vspace{-0.3cm}
\caption{Far-field UPW beam patterns of modular and collocated ULAs versus spatial frequency differences $\Delta_\theta$.}\label{movement}
  \vspace{-0.6cm}
\end{figure}

\begin{figure}
  \vspace{-0.6cm}
  \centering
  \subfigure[]{
    \label{1} 
    \includegraphics[scale=0.25]{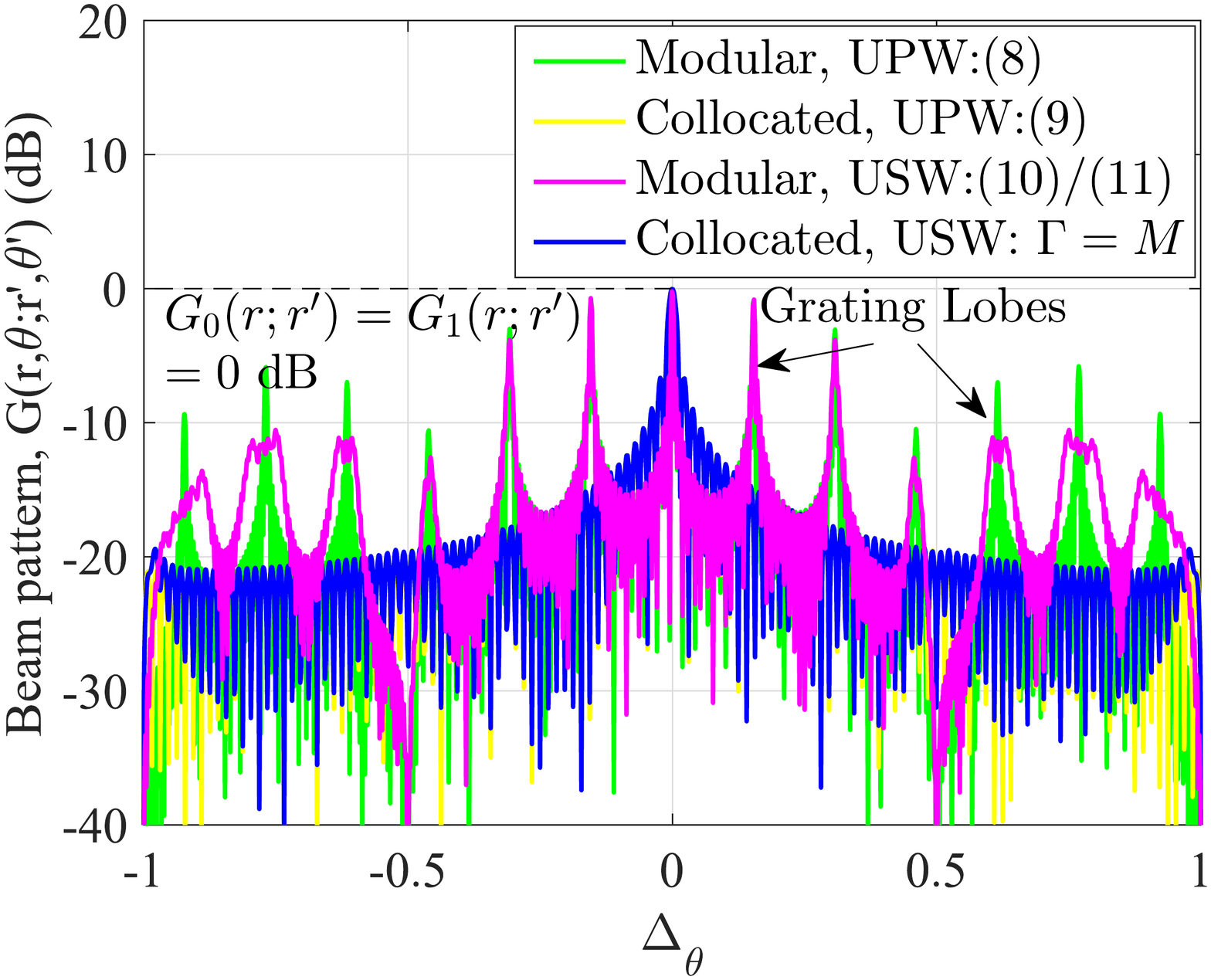}}
  \hspace{0in}
  \subfigure[]{
    \label{2}
    \includegraphics[scale=0.25]{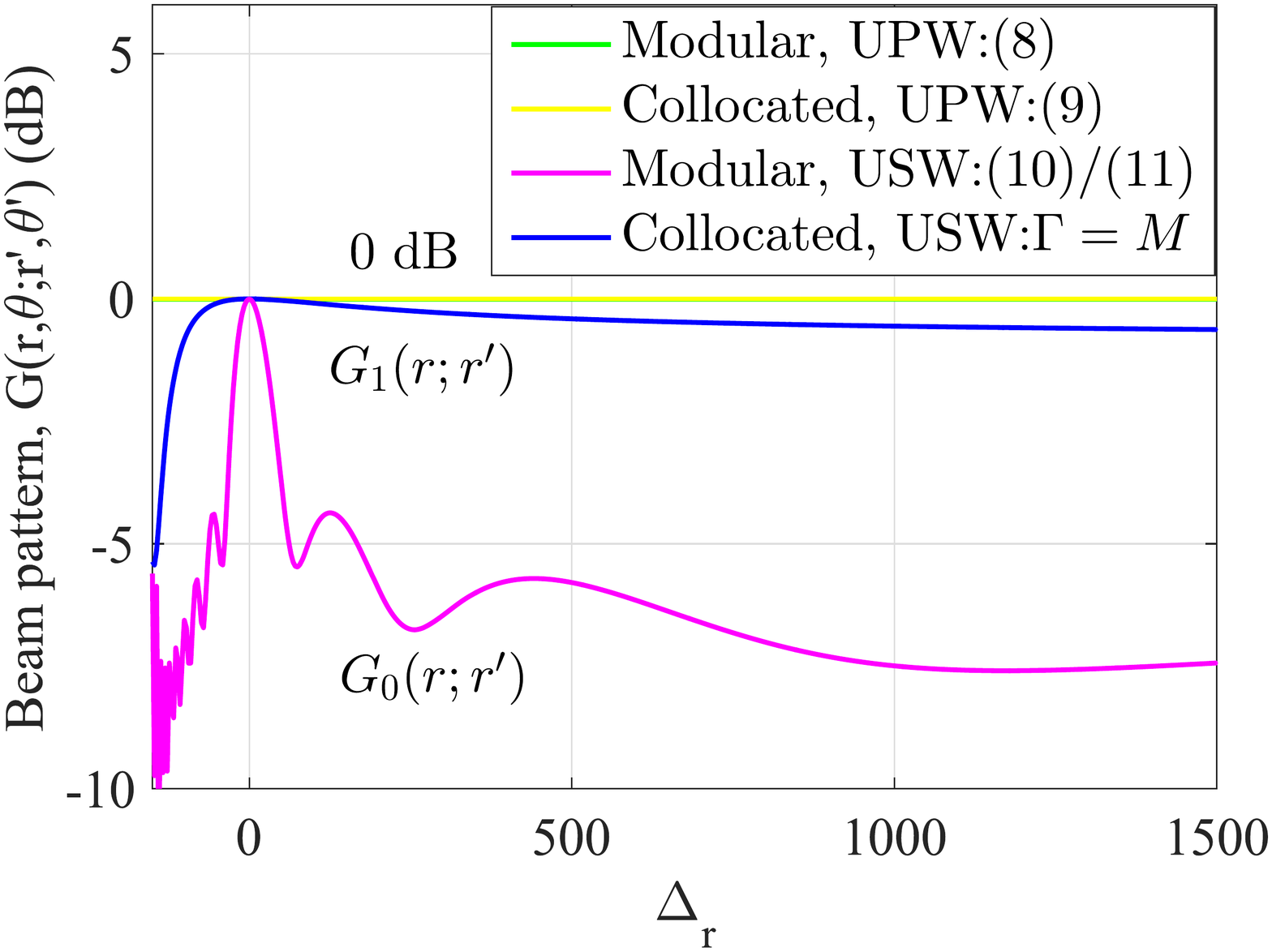}}
    \hspace{0in}
    \subfigure[]{
    \label{2}
    \includegraphics[scale=0.254]{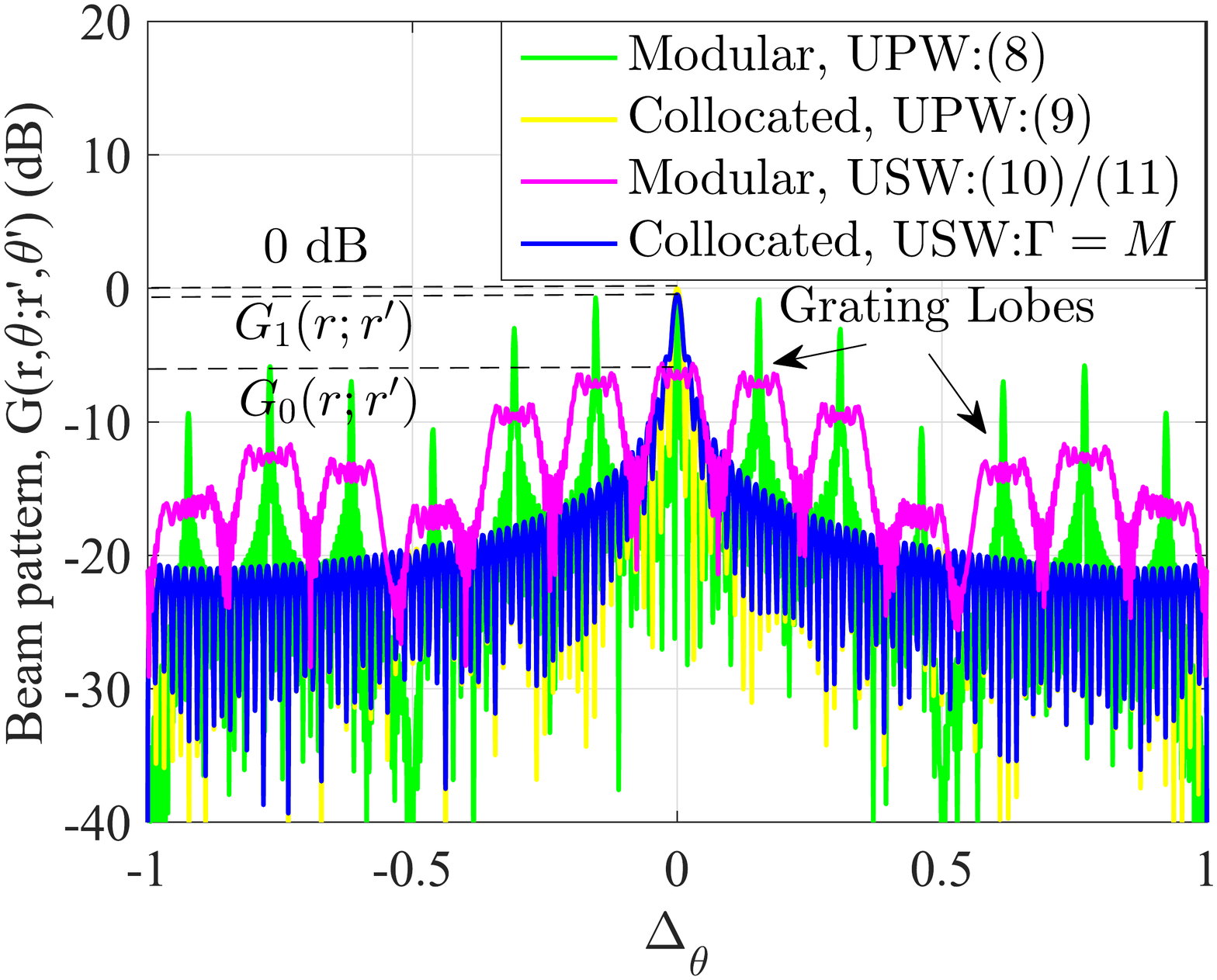}}
    \hspace{0in}
    \vspace{-0.3cm}
  \caption{Near-field USW beam focusing patterns of modular and collocated XL-ULAs: (a) versus spatial frequency differences $\Delta_\theta$, with $r=r'=200$ $\rm m$; (b) versus distance differences $\Delta_r$, with $\theta=\theta'=0$; (c) versus spatial frequency differences $\Delta_\theta$, by fixing $r=800$ $\rm m$ and $r'=200$ $\rm m$.}\label{pic1}
  \label{12}
  \vspace{-0.1cm}
\end{figure}

\section{Simulation results} \label{model}\vspace{-1pt}


In this section, numerical results are provided to illustrate and verify
the beam focusing patterns and grating lobes of mXL-ULA communications. For the following plots on beam patterns, the intended beam focusing point is $(r', \theta')=($200 \rm  m$, 0)$. The inter-element spacing is $d=\frac{\lambda}{2}=0.0628$ $\rm m$.

Fig. 3 plots the far-field UPW beam patterns of the collocated and modular ULAs with equal number of elements. For modular ULA, the number of modules is $N=4$, the number of antennas in each module is  $M=4$, and the modular size parameter is $\Gamma=13$.  It is observed that the main lobe beam width of modular array $\frac{2}{\Gamma N\bar{d}}$ is much smaller than its collocated counterpart $\frac{2}{M N\bar{d}}$. This illustrates that modular array can achieve the higher angular resolution than the collocated counterpart.
It is also observed that for modular array, grating lobes with a period of $\frac{1}{\Gamma \bar d}$ appear in the angular domain since $\Gamma\bar{d}>\frac{1}{2}$, which can cause angular ambiguity for communication or sensing. Fortunately, the term $\left|H_{M,\bar{d}}(\Delta_\theta)\right|$ in (8) can serve as an envelope of $G(\theta;\theta')$ to slightly suppress the grating lobes.

Fig. 4 plots the near-field USW beam focusing patterns of the collocated and modular array architectures, i.e., (a) versus spatial frequency differences, with $r=r'=200$ $\rm m$; (b) versus distance differences, with $\theta=\theta'=0$; (c) versus spatial frequency differences, by fixing $r=800$ $\rm m$ and $r'=200$ $\rm m$.
For modular XL-ULA, we set $N=32$, $M=4$, and $\Gamma=13$. It is found that for the given settings, the two USW beam focusing patterns of mXL-ULA in (10) and (11) match quite well. As shown in Fig. 4(a) and (b), compared with its collocated counterpart, the mXL-ULA under the USW model shows its superiority for improving the spatial resolution in both angular and distance dimensions, but at the cost of generating grating lobes. Moreover, as observed in Fig. 4(b) and (c), benefiting from its better distance resolution, modular architecture can better mitigate the IUI among users along the same direction as compared to its collocated counterpart, since $G_0(r;r')\le G_1(r;r')\le 1$ or $0$ $\rm dB$. Note that for the same spatial frequency, the UPW beam patterns of two considered architectures always keep a constant value of $0$ $\rm dB$, regardless of distance variations. It is also seen from Fig. 4(a) and (c) that due to the non-linear phase variations across array elements of mXL-ULA, the USW model shows a higher  grating lobe suppression capability than the UPW model.\par

\section{Conclusions}\label{g} \vspace{-1pt}

This paper studied the near-field modelling and beam focusing pattern analysis for mXL-ULA communications.
To simplify the NUSW model, we presented sub-array based USW models with different or common angles, under which
the near-field beam focusing patterns of mXL-ULA were developed. It was revealed that compared to the conventional collocated counterpart with equal number of array elements, mXL-ULA can significantly enhance the spatial resolution, but at the cost of generating undesired grating lobes. Moreover, due to
 the non-linear phase variations across
array elements, the near-field USW model for mXL-ULA exhibited a higher  grating lobe suppression capability. Numerical results further verified these important characteristics of beam focusing patterns and grating lobes for mXL-ULA.

\section*{Acknowledgement}
This work was supported by the National Key R$\&$D Program of China with Grant number 2019YFB1803400, and
the Fundamental Research Funds for the Central Universities of China under grant number 2242022k60004.

\appendices
\section{Proof of sub-array based USW model with common angle}
To derive the condition under which (5) can be simplified as (6), we rewrite ${\bf b}(\theta_n)$ in (5) in an equivalent form as
\begin{equation}\label{EQU-21}
\begin{split}
&\!{\bf b}(\theta_n)\!=\!
\!\left[e^{j\frac{2\pi}{\lambda}md\sin\theta}\!\right]_{m\in\mathcal M}\!\!\!\odot \! \!\left[e^{j\frac{2\pi}{\lambda}md\!(\sin\theta_n\!-\!\sin\theta)}\!\right]\!_{m\in\mathcal M}\!,\! \forall n\!\in\!\mathcal N.\!\\
\end{split}
\end{equation} \vspace{-0.1cm}\par
Obviously, (5) degenerates to (6) if the second term in (13) is negligible, i.e.,
\begin{equation}\label{EQU-13}
\begin{split}	
&\frac{2\pi}{\lambda}\left| md(\sin\theta_n-\sin\theta)\right|\leq \epsilon, \forall n\in\mathcal N, m\in\mathcal M,\\
\end{split}
\end{equation}
where $\epsilon\ll 1$.  To this end,  $\sin\theta_n$ is firstly approximated as
\begin{equation}\label{EQU-13}
\begin{split}	
\sin\theta_n&=\frac{r\sin\theta-y_n}{\sqrt{r^2-2r\sin\theta y_n+y_n^2}}\overset{(\rm a)}{\approx }\sin\theta-\frac{y_n}{r}, \forall n\in\mathcal N,\\
\end{split}
\end{equation}
where (a) holds if $r\gg |y_n|, \forall n\in\mathcal N$, which is equivalent to $r\gg \frac{D}{2}$. By choosing $10$ times as the criterion for ``much greater", we have the first condition
$r\ge5D$. By substituting (15) into (14) and letting  $\epsilon=\frac{\pi}{8}$, we have
\begin{equation}\label{EQU-13}
\begin{split}
\frac{2\pi}{\lambda}\left|md\frac{y_n}{r}\right|=\frac{2\pi}{\lambda}\left|\frac{m n\Gamma d^2}{r}\right|\le\frac{\pi}{8}, \forall n\in\mathcal N, m\in\mathcal M.\\
\end{split}
\end{equation} \par
Since $|m|\le\frac{M-1}{2}$, $|n|\le\frac{N-1}{2}$, and $N\gg1$, we have $r \ge \frac{4(M-1)(N-1)\Gamma d^2}{\lambda}\approx\frac{4(M-1)dD}{\lambda}=\frac{4SD}{\lambda}$. Therefore, we achieve the conditions for (6), i.e.,
$r \ge \max\left\{5D, \frac{4SD}{\lambda}\right\}$. It is observed that when $d=\frac{\lambda}{2}$, we have $\frac{4SD}{\lambda}=2(M-1)D$, larger than $5D$ under $M>3$. In addition, due to $N\!\gg \!1$ so that $D>2S$, $\frac{2S^2}{\lambda}<\max\left\{5D, \frac{4SD}{\lambda}\right\}<\frac{2D^2}{\lambda}$ is guaranteed. 

\section{Proof of Corollary 2}
To obtain closed-form expression for (11), we first consider the approximations $r_{n}\approx r-y_n\sin\theta+\frac{y_n^{2}\cos^{2}\theta}{2 r}$  and $r'_{n}\approx r'-y_n\sin\theta'+\frac{y_n^{2}\cos^{2}\theta'}{2 r'}$, $\forall n\in\mathcal N$, by using the second-order Taylor expansion, i.e.,
$\sqrt{1+x} \approx 1+\frac{1}{2} x-\frac{1}{8} x^{2}$ \cite{Gradshteyn2007, Dai2022}. This leads to
$\Delta_{r,n}=r_{n}-r'_{n}\approx \Delta_r-y_n\Delta_\theta+\frac{1}{2}y_n^{2}\delta_{r,\theta}$, $\forall n\in\mathcal N$. After that, we define the function $f(x)=e^{j\left(\nu_{r,\theta}N^2{x}^2+ \mu_{\theta} N{x}\right)}$  over the domain $\mathcal{A}=\left\{x|-\frac{N}{2}\epsilon\le x \le\frac{N}{2}\epsilon\right\}$. Since $\epsilon=\frac{1}{N} \ll 1$, we have $f(x)\approx f(n\epsilon)$, $\forall n\in\mathcal N$. Then, according to $\nu_{r,\theta}$, we ought to discuss three cases as follows:\par
1) When $\frac{\cos^2\theta}{r}<\frac{\cos^2\theta'}{r'}$, (11) can be rewritten as
\begin{equation}\label{EQU-9}
\begin{split}
&G(r,\theta;r',\theta')=\frac{1}{N}\left|\sum_{n\in\mathcal N} e^{j\left(\nu_{r,\theta}{n}^2+ \mu_{\theta} {n}\right)}\right|\left|H_{{M},\bar{d}}(\Delta_\theta)\right|\\
&\stackrel{(\rm a)}{\approx}\left|\int_{x\in\mathcal{A}} e^{j(\nu_{r,\theta}N^2x^2+ \mu_{\theta}Nx)}\!dx \right|\!\left|H_{{M},\bar{d}}(\Delta_\theta)\!\right|\\
&\stackrel{(\rm b)}{=}\!\frac{\left|\!F\left(\sqrt{\nu_{r,\theta}}\frac{N}{2}\!\!+\!\!\frac{\mu_{\theta}}{2\sqrt{\nu_{r,\theta}}}\right)\!+\!F\left(\!\sqrt{\nu_{r,\theta}}\frac{N}{2}\!\!-\!\frac{\mu_{\theta}}{2\sqrt{\nu_{r,\theta}}}\right)\!\right|\!}{\sqrt{\nu_{r,\theta}}N}
\!\left|H_{{M},\bar{d}}(\!\Delta_\theta)\!\right|.
\end{split}
\end{equation}
where (a) holds due to  the integral approximation, i.e., $\sum_{n\in\mathcal N}f(n\epsilon)\epsilon\approx\int_{x\in\mathcal{A}}f(x)dx$ \cite{Zeng2021, Dai2022}; (b) is derived by using Euler formula, i.e., $e^{jx}=\cos x+j\sin x$,
integral results, i.e., $\int \sin \left(a x^{2}+2 b x+c\right) \mathrm{d} x=\frac{1}{\sqrt{a}}
\left\{\cos \left(\frac{a c-b^{2}}{a}\right) S\left(\frac{a x+b}{\sqrt{a}}\right)+\sin \left(\frac{a c-b^{2}}{a}\right) C\left(\frac{a x+b}{\sqrt{a}}\right)\right\}$, $a>0$, and
$\int \cos \left(a x^{2}+2 b x+c\right) \mathrm{d} x=\frac{1}{\sqrt{a}}
\left\{\cos \left(\frac{a c-b^{2}}{a}\right) C\left(\frac{a x+b}{\sqrt{a}}\right)-\sin \left(\frac{a c-b^{2}}{a}\right) S\left(\frac{a x+b}{\sqrt{a}}\right)\right\}$, $a>0$, where $C(x)= \int_{0}^{x} \cos t^{2} \mathrm{~d} t$ and $S(x)=\int_{0}^{x} \sin t^{2} \mathrm{~d} t$ are Fresnel integrals \cite{Gradshteyn2007}, as well as  $F(x)=C(x)+jS(x)$.\par
2) When $\frac{\cos^2\theta}{r}=\frac{\cos^2\theta'}{r'}$, (11) can be expressed as
$G(r,\theta;r',\theta')=
\left|H_{N,\Gamma\bar{d}}(\Delta_\theta)\right|\left|H_{M,\bar{d}}(\Delta_\theta)\right|$, same to the UPW result in (8).

3) When $\frac{\cos^2\theta}{r}>\frac{\cos^2\theta'}{r'}$, similar to (17) and due to space limit, (11) can be easily formulated as
$G(r,\theta;r',\theta')= \frac{\left|F\!\left(\!\sqrt{-\nu_{r,\theta}}\frac{N}{2}+\frac{\mu_{\theta}}{2\sqrt{-\nu_{r,\theta}}}\!\right)+F\left(\!\sqrt{-\nu_{r,\theta}}\!\frac{N}{2}-\frac{\mu_{\theta}}{2\sqrt{-\nu_{r,\theta}}}\!\right)\!\right|}{\sqrt{-\nu_{r,\theta}}N}
\left|\!H_{M,\bar{d}}(\Delta_\theta\!)\right|$.\par
This thus completes the proof of Corollary 2.

\bibliographystyle{IEEEtran}
\end{document}